\documentclass[
aps,
prx,
reprint,
superscriptaddress,
floatfix,
longbibliography
]{revtex4-2}

\usepackage[T1]{fontenc}
\usepackage[utf8]{inputenc}
\usepackage[english]{babel}
\usepackage{csquotes}
\usepackage{amsmath}
\usepackage{amsmath,amssymb,amsfonts}
\usepackage{mathtools}
\usepackage{bm}
\usepackage{physics}

\usepackage{graphicx}
\usepackage[dvipsnames]{xcolor}
\usepackage[font=footnotesize,labelfont=bf]{caption}
\usepackage{subcaption}
\usepackage{xr-hyper}
\usepackage[colorlinks=true,linkcolor=MidnightBlue,citecolor=MidnightBlue,urlcolor=MidnightBlue]{hyperref}
\usepackage[nameinlink,noabbrev]{cleveref}
\begin{document}

    \title{Toward Quantum Utility in Correlated Topological Matter: Variational Preparation of Fractional Quantum Hall Manifolds}
    	
	\author{Sergio F. Expósito}
	\affiliation{Donostia International Physics Center (DIPC), Paseo Manuel Lardizabal 4, 20018 Donostia/San Sebastián, Spain}
	\affiliation{Fisika Aplikatua Saila, Gipuzkoako Ingeniaritza Eskola, University of the Basque Country (UPV/EHU), Europa Plaza 1, 20018 Donostia/San Sebastián, Spain}

	\author{Unai Aseginolaza}
	\affiliation{Basic Sciences Department, Faculty of Engineering, Mondragon Unibertsitatea, 20500 Arrasate, Spain}
	
	\author{Raúl Guerrero-Avilés}
	\affiliation{TECNALIA, Basque Research and Technology Alliance (BRTA), Astondo Bidea, Edificio 700, E-48160 Derio, Bizkaia, Spain}
	
	\author{Joaquim Jornet-Somoza}
	\affiliation{Servicios Generales a la Investigaci\'on (SGIker), University of the Basque Country (EHU), Avenida de Tolosa 72, 20018 Donostia/San Sebastián, Spain}
	
	\author{Francisco Guinea}
	\affiliation{Donostia International Physics Center (DIPC), Paseo Manuel Lardizabal 4, 20018 Donostia/San Sebastián, Spain}
	\affiliation{IMDEA Nanoscience, Faraday 9, 28049 Madrid, Spain}

    	\author{Juan Borge}
	\affiliation{Fisika Aplikatua Saila, Gipuzkoako Ingeniaritza Eskola, University of the Basque Country (EHU), Europa Plaza 1, 20018 Donostia/San Sebastián, Spain}

    \affiliation{Centro de F\'isica de Materiales (CFM-MPC), CSIC-UPV/EHU, Donostia 20018, Spain}

	\date{\today}

    \begin{abstract}
We investigate the use of variational quantum algorithms to prepare and characterize fractional quantum Hall states on near-term quantum processors. Focusing on the $\nu=1/3$ Laughlin phase described by the $V_1$ Haldane pseudopotential, we formulate the lowest-Landau-level problem in second quantization, and implement particle-number-preserving variational circuits combined with the variational quantum eigensolver (VQE) and variational quantum deflation (VQD). We benchmark the approach in two complementary geometries: Haldane sphere and torus shape. On the Haldane sphere, the target state is a unique zero-energy Laughlin ground state, providing a controlled test of the variational workflow and of excited-state reconstruction. On the torus, the problem retains the genuinely two-dimensional periodic character of the quantum Hall liquid and exhibits the threefold topological ground-state degeneracy expected for the $\nu=1/3$ fractional filling factor. This feature makes the torus a more demanding benchmark than the quasi-one-dimensional cylinder or thin-torus limits commonly exploited in state-preparation quantum protocols. We benchmark the hardware-optimized variational states against exact diagonalization using energy estimates, error-mitigated observables, and subspace-containment diagnostics. Our results show that hybrid quantum algorithms can approximately reconstruct the low-energy structure of small fractional quantum Hall systems, including the topological ground-state manifold on the torus. Beyond serving as a benchmark for quantum hardware, this geometry-resolved approach provides a route toward quantum simulations of fractional Chern insulators and strongly correlated topological phases in realistic two-dimensional materials.

    \end{abstract}

    \maketitle

    \section{Introduction}

Strongly interacting quantum systems are among the most fascinating regimes in condensed matter physics. They host phenomena with no simple single-particle counterpart, including Mott physics \cite{Mott1949,Hubbard1963,Imada1998}, quantum magnetism \cite{Anderson1973,Anderson1987,Lee2006}, unconventional superconductivity \cite{Bednorz1986,Scalapino2012}, topological order and fractionalization \cite{Wen1990,Wen1991,Senthil2004}, and the fractional quantum Hall (FQHE) \cite{Tsui1982,Laughlin1983,Stormer1999}.
In these systems, the relevant physics emerges from collective many-body correlations, making them both conceptually rich and computationally challenging. Indeed, the dimension of the many-body Hilbert space grows exponentially with the degrees of freedom of the system, so exact classical simulations rapidly become intractable even for moderately sized systems. This exponential complexity was one of the original motivations for quantum simulation, as pointed out by Feynman \cite{Feynman1982} and later formalized in the context of universal quantum simulators \cite{Lloyd1996}. Quantum computers therefore provide a natural framework for addressing strongly correlated matter, because quantum states are represented directly in a quantum-mechanical Hilbert space.

At the same time, the capabilities of noisy intermediate-scale quantum (NISQ) processors have grown rapidly over the last decade. 
Quantum processors platforms, such as superconducting, trapped-ion, neutral-atom, and photonic platforms, have demonstrated increasingly large and controllable quantum registers, improved gate fidelities, mid-circuit measurement and reset, tunable connectivity, and more sophisticated error-mitigation and benchmarking protocols \cite{Barends2014,Debnath2016,Bernien2017,Arute2019,Wright2019,Jurcevic2021,Noel2022,Kim2023,Bluvstein2024}. Although present devices are still affected by noise and do not yet provide fully fault-tolerant quantum computation, they already allow controlled experiments with many-body quantum dynamics beyond the reach of simple few-qubit demonstrations \cite{Preskill2018,Endo2021}. This progress has stimulated a growing effort to use quantum hardware as a tool for condensed matter physics, including the simulation of spin models, lattice gauge theories, fermionic Hamiltonians, Hubbard-type systems, topological models, and real-time dynamics of interacting phases \cite{Georgescu2014,Altman2021,Bauer2020,Stanisic2022,Mi2021,Satzinger2021,Chen2023,Kechedzhi2024}. In this context, strongly correlated and topological systems provide particularly demanding benchmarks because they combine large Hilbert spaces, entanglement, degeneracies, and symmetry constraints, while also offering physically meaningful observables with which to assess the performance of quantum simulation workflows.

A paradigmatic example of such strongly correlated physics is the FQHE. Following the discovery of the integer quantum Hall effect, where the Hall conductance is quantized in integer multiples of ($e^2/h$) \cite{Klitzing1980}, the observation of plateaux at fractional filling factors revealed a qualitatively different regime in which electron-electron interactions are essential \cite{Tsui1982}. At partial filling of a highly degenerate Landau level, the kinetic energy is quenched and the Coulomb interaction drives the formation of incompressible quantum liquids with fractionally charged excitations and topological order. Laughlin's variational wave function for the $\nu=1/3$ state provided the first microscopic description of this phenomenon \cite{Laughlin1983}, while Haldane's pseudopotential construction and spherical geometry established a powerful framework for identifying model Hamiltonians and finite-size spectra of fractional quantum Hall phases \cite{Haldane1983}. The FQHE has since become one of the clearest examples of topological order, anyonic quasiparticles, and interaction-driven many-body topology \cite{Arovas1984,Wen1990,Nayak2008,Stormer1999}.

Beyond continuum Landau levels, closely related phases can also emerge in lattice systems with nearly flat topological bands. These systems, known as fractional Chern insulators (FCI), can be understood as lattice analogues of fractional quantum Hall states, where interactions partially fill a Chern band and stabilize topologically ordered many-body phases without the need for an external magnetic field \cite{Tang2011,Sun2011,Neupert2011,Regnault2011}. This connection makes FCI especially attractive for quantum materials and moiré platforms, but also particularly difficult to address theoretically: they combine strong interactions, nontrivial band geometry, lattice effects, and large many-body Hilbert spaces \cite{Parameswaran2013,Bergholtz2013,Spanton2018,Cai2023}. In this sense, fractional quantum Hall systems provide both a canonical benchmark and a controlled starting point for developing quantum-computing workflows aimed at more general strongly correlated topological phases, including FCI.

The possibility of addressing fractional quantum Hall physics with quantum computers has recently attracted increasing attention. A key observation is that the two-dimensional problem can be represented in terms of a one-dimensional chain of guiding-center orbitals after projection onto the lowest Landau level (LLL). This mapping is particularly explicit in Landau gauge and becomes especially useful in cylindrical or thin-torus geometries, where fractional quantum Hall states acquire a quasi-one-dimensional structure while retaining signatures of their topological origin \cite{RezayiHaldane1994,Seidel2005,SeidelLee2007,BernevigHaldane2008,Lauchli2010,Bergholtz2013}. 
Recent studies on quantum algorithms inspired by one-dimensional structures have proposed quantum-circuits which are able to recreate Laughlin-type states \cite{Rahmani2020}. 
These states allow to construct linear depth circuits based on nearest-neighbors gates and quasi-one-dimensional representation of the $\nu = 1/3$ state.
More recently, related thin-cylinder and parent-Hamiltonian ideas have been used to formulate quantum-simulation protocols for geometric excitations and non-Abelian fractional quantum Hall states \cite{Kirmani2022,Voinea2024}. A major experimental step in this direction was the realization of a fermionic $\nu=1/3$ Laughlin state on a trapped-ion quantum processor using a symmetry-preserving Hamiltonian variational ansatz, demonstrating that digital quantum hardware can prepare and characterize strongly correlated topological states beyond purely bosonic or analog settings \cite{Shen2026}. These advances show that fractional quantum Hall states provide a natural benchmark for quantum processors: they combine a compact second-quantized representation, highly nontrivial many-body correlations, and clear diagnostic observables such as density correlations, entanglement signatures, edge structure, and topological degeneracy.

In this work, we address fractional quantum Hall states using hybrid variational quantum algorithms.
More specifically, we combine particle-number-preserving parametrized quantum circuits with 
Variational Quantum Eigensolver (VQE) and Variational Quantum Deflation (VQD) algorithms \cite{Peruzzo2014,Higgott2019,Cerezo2021} to reconstruct not only ground states, but also low-lying excited states and degenerate ground state manifolds. We focus on the $\nu=1/3$ Laughlin phase described by the $V_1$ Haldane pseudopotential \cite{Haldane1983}, which provides an exactly solvable parent Hamiltonian and therefore a stringent benchmark for quantum-computing workflows. The use of VQE allows us to approximate the lowest-energy state within a fixed particle-number sector, while VQD enables a sequential reconstruction of excited states or, in the torus geometry, of the different states belonging to the topological ground-state manifold.

A central aspect of our approach is the explicit comparison between two standard geometries of fractional quantum Hall physics: the sphere and the torus. On the sphere, the $\nu=1/3$ Laughlin state appears as a unique zero-energy ground state at flux $N_\phi=3(N_e-1)$, (where $N_{\phi}$ is the number of magnetic fluxes and $N_e$ the number of electrons) , with rotational symmetry providing angular-momentum quantum numbers and a clear diagnostic of the incompressible liquid \cite{Haldane1983}. On the torus, by contrast, there is no curvature and no shift \cite{WenZee1992}, so that $N_\phi=3N_e$, and the emergence of a threefold ground-state degeneracy associated with the genus-one topology defines the topological feature \cite{Haldane1985,WenNiu1990}. Recovering this manifold is a more demanding task than targeting a single state, because the algorithm must identify an entire low-energy subspace rather than an isolated eigenvector. 

In addition to the energy, we therefore compute geometry-resolved symmetry diagnostics. For the sphere, we evaluate the expectation value of the total angular momentum operator, $\langle \hat{L}^2\rangle$, which allows us to verify whether the variational states belong to the expected angular-momentum sector.
In particular, the Laughlin ground state on the sphere should appear as a $L=0$ singlet state. For the torus, where rotational symmetry is absent, we instead resolve the states according to the magnetic-translation momentum quantum number $K$ \cite{Ortiz2013}. This provides the natural analogue of the angular-momentum classification and allows us to identify the momentum sectors associated with the threefold topological ground-state manifold. These observables provide an important check beyond the variational energy, since a low-energy state may still have an incorrect symmetry character or may leak into neighboring sectors.
We also complement these symmetry diagnostics with a subspace-containment analysis, which tests whether the hardware-optimized variational states have large projection onto the exact low-energy eigenspaces and therefore distinguishes genuine wave-function reconstruction from agreement in energy alone

The torus formulation is particularly important in the context of previous quantum-computing approaches to fractional quantum Hall physics. Much of the recent literature exploits cylindrical or thin-torus limits, where the Landau-gauge orbital basis maps the problem onto a quasi-one-dimensional chain and leads to efficient circuit constructions for Laughlin-type states \cite{RezayiHaldane1994,Seidel2005,SeidelLee2007,BernevigHaldane2008,Rahmani2020,Shen2026}. 
These geometries are extremely valuable because they make the guiding-center structure transparent and reduce the complexity of state preparation. However, they also emphasize a quasi-one-dimensional limit of the problem. By working directly on the torus with periodic boundary conditions in both directions, we retain the genuinely two-dimensional character of the fractional quantum Hall liquid, including magnetic-translation momentum sectors and topological ground-state degeneracy. This provides a closer benchmark for future applications to FCI and quantum materials, where the relevant phases arise in two-dimensional lattice bands rather than in an effectively one-dimensional cylinder. In this sense, the torus geometry offers a natural bridge between continuum fractional quantum Hall physics and interaction-driven topological phases in realistic flat-band, moiré and Chern-band systems \cite{Tang2011,Sun2011,Neupert2011,Regnault2011,Parameswaran2013,Bergholtz2013}.

The paper is structured as follows. In Sec.~\ref{sec:model}, we introduce the fractional quantum Hall Hamiltonian projected onto the LLL and formulate the $V_1$ Haldane pseudopotential in second quantization. We then describe its implementation in the two geometries considered in this work: the Haldane sphere and the torus. In Sec.~\ref{sec:quantum_algorithm} we explain the quantum algorithms used to realize all the calculations. In Sec.~\ref{sec:results}, we present the variational quantum-computing calculations. 
We first analyze the torus geometry, where VQD is used to reconstruct the three-dimensional ground-state manifold and the first excited state, and then discuss the spherical geometry, including the VQE ground state and the VQD first excited state. 
We benchmark the hardware-optimized variational states against exact diagonalization using energies, error-mitigated estimates, subspace-containment diagnostics, and geometry-dependent symmetry observables. Finally, in Sec.~\ref{sec:conclusions}, we summarize the main results and discuss possible extensions. 
    
    \section{Model}
    \label{sec:model}
In this section we define the fractional quantum Hall Hamiltonians used throughout the work and fix the notation for their implementation in the geometries considered below. We follow the standard second-quantized formulation of lowest-Landau-level projected interactions \cite{Ortiz2013}, in which the kinetic energy is quenched and the many-body problem is entirely governed by the projected two-body interaction. In this language, the orbital index labels guiding-center degrees of freedom, and the quantum Hall problem can be written as an effective fermionic lattice Hamiltonian with interaction matrix elements constrained by the symmetries of the geometry. In particular, we focus on the fermionic $V_1$ Haldane pseudopotential, or Trugman--Kivelson Hamiltonian, whose zero-energy ground states at $\nu=1/3$ are the Laughlin states. We first introduce the general LLL projection and the pseudopotential Hamiltonian, and then specialize it to the two geometries used in this work. The torus realizes a genuinely two-dimensional periodic system with magnetic-translation symmetry and a threefold topological ground-state manifold, while the sphere provides a rotational invariant setting with a unique $L=0$ Laughlin ground state.
The basic Hamiltonian of the fractional quantum Hall 
problem is given by
\begin{equation}
\hat{H}_{\mathrm{QH}}=\hat{P}_{\mathrm{LLL}}
\hat{H}_{\mathrm{int}}
\hat{P}_{\mathrm{LLL}} .
\label{eq:H_QH_projection}
\end{equation}
This Hamiltonian describes a two-dimensional gas of $N_e$ spin-polarized electrons in a strong perpendicular magnetic field, after projection onto the lowest Landau level. Working only on the LLL is accurate when the cyclotron gap is much larger than the interaction energy scale, so that Landau-level mixing can be neglected. In this limit, the kinetic energy is fixed by the Landau-level index and becomes an irrelevant constant. The many-body physics is therefore governed entirely by the electron-electron interaction projected onto the LLL,
\begin{equation}
\hat{H}_{\mathrm{QH}}=
\frac{1}{2}
\int d^2 r d^2 r'
\hat{\psi}^{\dagger}(\mathbf{r})
\hat{\psi}^{\dagger}(\mathbf{r}')
V(\mathbf{r}-\mathbf{r}')
\hat{\psi}(\mathbf{r}')
\hat{\psi}(\mathbf{r}) .
\label{eq:H_QH_continuum}
\end{equation}

Here, $V(\mathbf{r}-\mathbf{r}')$ is the two-body interaction and $\hat{\psi}(\mathbf{r})$ is the field operator projected onto the LLL. In order to map the problem to a second-quantized orbital representation, the field operator can be expanded in the basis of LLL orbitals as follows,
\begin{equation}
\hat{\psi}(\mathbf{r})=
\sum_{j=0}^{N_L-1}
\phi_j(\mathbf{r})\hat{c}_j,
\label{eq:field_LLL}
\end{equation}
where $N_L$ is the number of orbitals in the LLL, $\phi_j$ is the $j$-th orbital on the LLL and $\hat{c}_j$ is the annihilation operator acting on the $j$-th orbital. The orbital index $j$ labels guiding-center degrees of freedom rather than ordinary lattice positions. Nevertheless, once the LLL projection has been performed, the fractional quantum Hall problem can be viewed as an effective fermionic lattice model in this guiding-center basis.

In the following, we specialize this projected Hamiltonian to the two geometries considered in this work. We first discuss the torus, where the relevant symmetries are magnetic translations and the Laughlin phase is characterized by a topological ground-state threefold degeneracy. We then discuss the sphere case, where rotational symmetry provides an alternative classification in terms of total angular momentum.

\subsection{Torus geometry}
\label{subsec:torus_geometry}

We consider a rectangular torus of lengths $L_x$ and $L_y$, pierced by $N_{\phi}$ magnetic flux quanta. Its area satisfies
\begin{equation}
L_xL_y=2\pi \ell_B^2 N_{\phi},
\label{eq:torus_area_flux}
\end{equation}
where $\ell_B$ is the magnetic length. In the Landau gauge, the LLL is spanned by $N_{\phi}$ guiding-center orbitals labelled by $r=0,\ldots,N_{\phi}-1$. The torus closes the orbital chain by periodic boundary conditions, so that orbital labels are understood modulo $N_{\phi}$,
\begin{equation}
\hat{c}_r \equiv \hat{c}_{r+N_{\phi}} .
\label{eq:torus_orbital_periodicity}
\end{equation}
This periodic identification is the key difference with respect to cylindrical geometries. On a cylinder, the Landau-gauge orbitals form effectively quasi-one-dimensional chain with physical open edges, while on the torus both spatial directions are compact. 
As a result, the torus has no edge and no global rotational symmetry. The many-body states are therefore not classified by total angular momentum, but by magnetic-translation momentum sectors.

On the torus, the projected interaction conserves the guiding-center center-of-mass modulo $N_{\phi}$. For $N_L=N_{\phi}$ orbitals and after performing the integrals in Eq.~\ref{eq:H_QH_continuum} for the $V_1$ Haldane pseudopotential of the torus we can write the Hamiltonian in its factorized form 
\begin{equation}
\label{eq:H_torus}
\hat{H}_{V_1}^{\mathrm{torus}}
=
\sum_j
\sum_{k,l}
\eta_k\eta_l
\hat{c}^\dagger_{j+k}
\hat{c}^\dagger_{j-k}
\hat{c}_{j-l}
\hat{c}_{j+l},
\end{equation}
where all orbital indices are understood modulo $N_L$. The sums run over values of $j$, $k$, and $l$ such that $j\pm k$ and $j\pm l$ are valid orbital labels modulo $L$. As in the center-of-mass notation introduced above, $j$, $k$, and $l$ may be either all integers or all half-odd integers. This ensures that the creation and annihilation operators always act on integer-labelled orbitals.

The periodic boundary conditions are also encoded in the interaction coefficients. They satisfy
\begin{equation}
\eta_{k+N_L}=\eta_k ,
\label{eq:eta_periodicity}
\end{equation}
and, for the torus $V_1$ pseudopotential, they are obtained by periodizing the cylinder coefficients,
\begin{equation}
\eta_k=
2\left(\frac{8}{\pi}\right)^{1/4}
\kappa^{3/2}
\sum_{s\in\mathbb{Z}}
(k+sN_L)
e^{-\kappa^2(k+sN_L)^2},
\label{eq:eta_torus}
\end{equation}
where $\kappa=2\pi/L_y$ in units of the magnetic length. Eq. ~\ref{eq:eta_torus} is the torus version of the short-range potential of the form $V(\mathbf{r})
\propto
-\nabla^2\delta^{(2)}(\mathbf{r}) $ in the guiding-center basis \cite{Ortiz2013}. For more details of this derivation see the Supplementary Material.

On the torus, continuous rotations are not a symmetry of the finite geometry. The relevant conserved quantities are instead associated with magnetic translations. In the Landau-gauge occupation basis, one component of the many-body magnetic momentum is generated by 
\begin{equation}
\hat{T}_y=
\exp\left[
\frac{2\pi i}{N_L}
\sum_{j=0}^{N_{\phi}-1}
j\hat{n}_j
\right],
\label{eq:Ty_operator}
\end{equation}
with
\begin{equation}
\hat{n}_j=\hat{c}_j^\dagger \hat{c}_j .
\label{eq:number_operator}
\end{equation}
For an occupation configuration $|n_0 n_1 \cdots n_{N_L-1}\rangle$, this operator gives
\begin{equation}
\hat{T}_y
|n_0 n_1 \cdots n_{N_L-1}\rangle=
e^{2\pi i K/N_L}
|n_0n_1\cdots n_{N_{L}-1}\rangle ,
\label{eq:Ty_eigenvalue}
\end{equation}
where
\begin{equation}
K=
\sum_{j=0}^{N_{L}-1}
j\hat{n}_j
\mod N_{L}.
\label{eq:torus_K}
\end{equation}
Therefore, the Hamiltonian in Eq.~\eqref{eq:H_torus} decomposes into independent magnetic-momentum sectors labelled by $K$. In the numerical calculations below, this quantum number provides the torus analogue of the angular-momentum diagnosis used on the sphere.

The torus has no curvature and therefore no spherical shift \cite{WenZee1992}. For the Laughlin sequence at filling $\nu=1/q$, the flux-particle relation is
\begin{equation}
N_L=qN_e .
\label{eq:torus_flux_relation}
\end{equation}
Thus, for the fermionic $\nu=1/3$ state considered here, $N_L=3N_e$. A defining feature of the torus geometry is the corresponding topological ground-state degeneracy: the $V_1$ parent Hamiltonian supports three zero-energy Laughlin ground states, separated from higher-energy states by an incompressibility gap \cite{HaldaneRezayi1985,WenNiu1990}.

\subsection{Haldane sphere}

We now consider the same $V_1$ pseudopotential problem on the Haldane sphere. 
In this spherical geometry introduced by Haldane \cite{Haldane1983}, the electrons move on the surface of a sphere pierced by a magnetic monopole at its center. We denote the monopole strength by $S$, so that the total number of magnetic flux quanta through the surface is
\begin{equation}
N_{\phi}=2S .
\label{eq}
\end{equation}
The single-particle states are monopole harmonics. In the LLL, their angular momentum is fixed by the monopole strength, $\ell=S$. Therefore, the LLL contains
\begin{equation}
N_{L}=2\ell+1=2S+1=N_{\phi}+1
\label{eq}
\end{equation}
single-particle orbitals, with angular-momentum projections $\mu=-S,\ldots,S$. Thus, the same quantity $S$ fixes both the magnetic flux and the angular momentum of the LLL multiplet.

Because the sphere preserves rotational symmetry, two-particle states in the LLL can be classified by their total pair angular momentum $J$ and its projection $M$. For two particles each carrying single-particle angular momentum $S$, the relative angular momentum $m$ is related to the total pair angular momentum by
\begin{equation}
m=2S-J .
\label{eq}
\end{equation}
The $V_1$ Haldane pseudopotential penalizes pairs with relative angular momentum $m=1$ \cite{Haldane1983}. On the sphere, this corresponds to the pair angular-momentum channel
\begin{equation}
J=2S-1 .
\label{eq}
\end{equation}

The rotational symmetry of the sphere allows the interaction to be written directly in terms of pair operators with well-defined total angular momentum. Since the $V_1$ pseudopotential selects the channel $J=2S-1$, we define the corresponding pair-annihilation operator as
\begin{equation}
T_M
=
\sum_{\mu_1,\mu_2=-S}^{S}
C^{2S-1,M}_{S\mu_1,S\mu_2}
c_{\mu_2}c_{\mu_1},\label{eq:sphere_pair_annihilation}
\end{equation}
where $c_\mu$ annihilates an electron in the LLL orbital with angular-momentum projection $\mu$, and $C^{2S-1,M}_{S\mu_1,S\mu_2}$ is the Clebsch--Gordan coefficient coupling two single-particle angular momenta $S_{\mu_1}$, $S_{\mu_1}$, into total pair angular momentum $J=2S-1$ and projection $M$. These coefficients enforce the angular-momentum selection rules of the spherical geometry. In particular, only pairs with total projection $M=\mu_1+\mu_2$ contribute.

The spherical $V_1$ Hamiltonian is then obtained by summing the corresponding positive semidefinite pair projectors over all allowed values of $M$,
\begin{equation}
\hat{H}_{V_1}^{\mathrm{sphere}}
=
\frac{V_1}{2}
\sum_{M=-(2S-1)}^{2S-1}
T_M^\dagger T_M .
\label{eq:H_sphere}
\end{equation}
This form makes explicit that the interaction penalizes only the two-particle component with relative angular momentum $m=1$. The factor $1/2$ compensates for the unrestricted sum over the two single-particle indices in the definition of $T_M$.

For the fermionic Laughlin state at filling $\nu=1/3$, the flux-particle relation on the sphere is
\begin{equation}
N_\phi=3(N_e-1).
\label{eq:sphere_flux_laughlin}
\end{equation}

For later use as a symmetry diagnostic, we also define the total many-body angular-momentum operator in second quantization. In the spherical LLL basis, the single-particle orbitals are labelled by the angular-momentum projection $\mu=-S,\ldots,S$. The many-body angular-momentum operators are
\begin{equation}
\hat{L}_z=
\sum_{\mu=-S}^{S}
\mu c_\mu^\dagger c_\mu ,
\label{eq:Lz_second_quantized}
\end{equation}
and
\begin{equation}
\hat{L}_+=
\sum_{\mu=-S}^{S-1}
\sqrt{(S-\mu)(S+\mu+1)}
c_{\mu+1}^\dagger c_\mu ,
\qquad
\hat{L}_-=\hat{L}_+^\dagger .
\label{eq:Lpm_second_quantized}
\end{equation}
The total angular momentum is then computed from
\begin{equation}
\hat{L}^2=
\hat{L}_z^2
+
\frac{1}{2}
\left(
\hat{L}_+\hat{L}_-
+
\hat{L}_-\hat{L}_+
\right).
\label{eq:L2_second_quantized}
\end{equation}
At the Laughlin flux, the spherical $V_1$ Hamiltonian in equation~\ref{eq} has a unique zero-energy ground state. Since $\hat{L}^2$ commutes with the spherical Hamiltonian, it provides a direct symmetry diagnostic for the variational states. The Laughlin ground state is expected to lie in the total angular momentum singlet sector, $L=0$, or equivalently $\langle \hat{L}^2\rangle=0$, identifying it as the spherical realization of the incompressible Laughlin fluid.

\begin{figure*}[t]
\centering
\includegraphics[width=\linewidth]{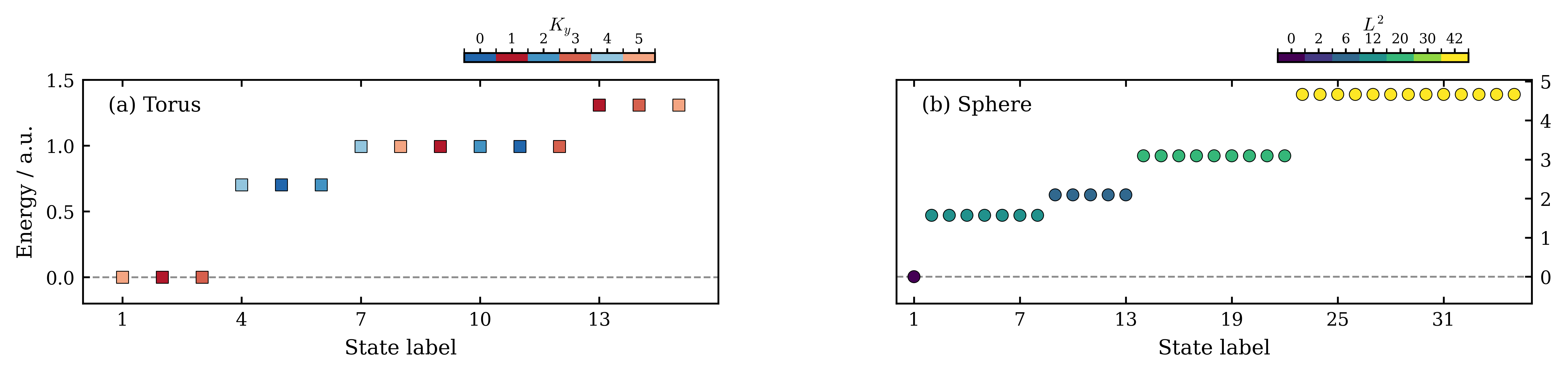}
\caption{Exact-diagonalization spectra of the \(V_1\) parent Hamiltonian for $N_L=6$, $N_e=2$ and $N_L=7$ and $N_e=3$ for the torus and sphere cases respectively. The torus spectrum is colored according to the magnetic-translation momentum sector \(K\), while the sphere spectrum is colored according to \(\langle \hat L^2\rangle\). The torus displays the expected threefold zero-energy $K=1,3,5$ ground-state degeneracy of the \(\nu=1/3\) Laughlin phase, whereas the sphere exhibits a unique zero-energy \(L=0\) ground state. }
\label{fig:ED_Comparison}
 \end{figure*}

The exact diagonalization spectra of the two described Hamiltonians for $N_L=7$ and $N_e=3$ sphere and $N_L=6$ and $N_e=2$ for the torus cases are shown in Fig.~\ref{fig:ED_Comparison}. The sphere displays a unique zero-energy ground state, with $\langle \hat{L}^2\rangle=0$, as expected for the Laughlin state at $\nu=1/3$. In contrast, the torus spectrum exhibits the characteristic threefold zero-energy ground-state manifold, with the states distributed across magnetic-momentum sectors labelled by $K=1,3,5$. These two spectra summarize the different symmetry structures of the geometries: rotational symmetry and angular momentum on the sphere, and magnetic translations and topological degeneracy on the torus.

\section{Quantum algorithm}
\label{sec:quantum_algorithm}

In this section we summarize the variational quantum computing workflow used to approximate the low-energy states of the FQHE Hamiltonians defined above. The goal is not to introduce a new quantum algorithm, but to adapt standard hybrid variational methods to the symmetry structure of the $V_1$ parent Hamiltonian in the torus and the sphere geometries. 
Concretely, we use VQE, where a parametrized circuit prepares a trial state, $|\psi(\boldsymbol{\theta})\rangle$, which is initialized in a reference state $|\psi_{0}\rangle$. The cost function calculates the expectation value of the energy, $\langle \psi(\boldsymbol{\theta})|\hat{H}|\psi(\boldsymbol{\theta})\rangle$. Both circuit and cost functions allows to classically minimize the parameter vector of the trial state and obtain the ground state energy and state by recovering the parameter vector \cite{Peruzzo2014}. Subsequently, we use VQD to obtain excited states and degenerate manifolds based on the penalization of the overlap with the previously optimized states. The penalization procedure is included in the cost function in a way that the subsequent optimizations converge to orthogonal low-energy states \cite{Higgott2019} .

To implement the calculation on real quantum hardware, the second-quantized Hamiltonians are mapped to Pauli operators acting on the actual qubits using the Jordan--Wigner transformation \cite{JordanWigner1928}. In this encoding, each LLL orbital is associated with one qubit, and the occupation number of that orbital is represented by the computational basis state of the corresponding qubit. The resulting qubit Hamiltonian is a weighted sum of Pauli strings, whose expectation values can be measured on a quantum processor.

The variational circuit, also known as ansatz, is chosen to preserve the number of particles. We use the particle-number-conserving ansatz introduced in Ref.~\cite{Gard2020}, adapted here to the LLL orbital basis of the FQHE problem. This is important because the target Laughlin states live in a fixed $N_e$ sector, and preserving particle number prevents the variational optimization from exploring unphysical sectors of the Hilbert space. Further description on the circuit structure, Pauli-string measurements, the classical optimization algorithm and the error-mitigation techniques are given in the Supplemental Material.

The workflow is carried out by combining real-hardware optimization with noiseless statevector diagnostics. The variational parameters are optimized on the IBM Heron-R2 quantum processor, specifically on the \texttt{IBM\_BasqueCountry} System Two. For selected parameter vectors obtained during the hardware optimization, we then perform noiseless statevector evaluations of the same circuits. These statevector calculations do not correspond to an independent statevector VQE optimization; rather, they provide an ideal reference for the states generated by the QPU-optimized parameters. This allows us to distinguish limitations of the variational ansatz and optimization landscape from errors induced by hardware noise and finite sampling. In both geometries, we evaluate the energy and the relevant symmetry diagnostics: the magnetic-momentum sector $K$ on the torus and the angular-momentum expectation value $\langle \hat{L}^2\rangle$ on the sphere.

The noiseless statevector evaluation also allows us to quantify the wave-function content of the states defined by the QPU-optimized parameters. For a variational state $|\psi(\boldsymbol{\theta})\rangle$ and an exact-diagonalization eigenspace $\mathcal H_\alpha$, we define the subspace containment
\begin{equation}
C_\alpha(\boldsymbol{\theta})
=\langle \psi(\boldsymbol{\theta})|\hat{P}_\alpha|\psi(\boldsymbol{\theta})\rangle ,
\label{eq:subspace_containment}
\end{equation}
where $\hat{P}_\alpha$ is the projector onto $\mathcal H_\alpha$. If ${|\phi_{\alpha,i}^{\mathrm{ED}}\rangle}$ is an orthonormal basis of that eigenspace, this becomes
\begin{equation}
C_\alpha(\boldsymbol{\theta})=
\sum_i
\left|
\langle
\phi_{\alpha,i}^{\mathrm{ED}}
|
\psi(\boldsymbol{\theta})
\rangle
\right|^2 .
\label{eq:subspace_containment_basis}
\end{equation}
For a nondegenerate eigenstate, this reduces to the usual squared overlap with the corresponding exact eigenvector. For a degenerate manifold, however, the projection onto the full eigenspace is the appropriate basis-independent quantity. This distinction is essential on the torus, where the $\nu=1/3$ Laughlin ground state is threefold degenerate and overlaps with individual exact eigenvectors depend on the basis chosen inside the degenerate subspace. The containment analysis therefore provides a wave-function diagnostic complementary to the energy: it tells us whether the QPU-optimized parameters define the correct low-energy state or manifold in the absence of hardware noise. Additional implementation details are given in the Supplemental Material.

\section{Results}
\label{sec:results}

We now apply the variational quantum-computing workflow described in Sec.~\ref{sec:quantum_algorithm} to the $V_1$ fractional quantum Hall Hamiltonians introduced in Sec.~\ref{sec:model}. We consider small systems for which exact diagonalization is available, allowing a direct benchmark of the variational energies, symmetry quantum numbers, and wave-function diagnostics. The variational parameters are optimized on the \texttt{IBM\_BasqueCountry} quantum processor, and the resulting parameter vectors are subsequently evaluated, again, with noiseless statevector simulations. This comparison allows us to separate the quality of the learned variational state from the effects of hardware noise, finite sampling, and measurement errors. We also use the exact-diagonalization spectra shown in Fig.~\ref{fig:ED_Comparison} as a reference for identifying the target ground-state and excited-state manifolds.

\subsection{Torus geometry}
\label{subsec:results_torus}

We first consider the torus geometry, which provides the most direct test of whether the variational workflow can reconstruct a degenerate topological ground-state manifold. We have computed the fermionic Laughlin state at $\nu=1/3$, with $N_L=6$ and $N_e=2$. In the finite system studied here, the $V_1$ Hamiltonian present a threefold degenerate ground state at zero energy, separated from the excited spectrum by a finite-size incompressibility gap. These three states belong to well-defined magnetic momentum sectors labelled by $K$, and their reconstruction therefore requires more than finding a single low-energy eigenvector. We use VQD sequentially to obtain the three states of the ground-state manifold and then the first excited state above it.

\begin{figure}
\centering
\includegraphics[width=0.9\columnwidth]{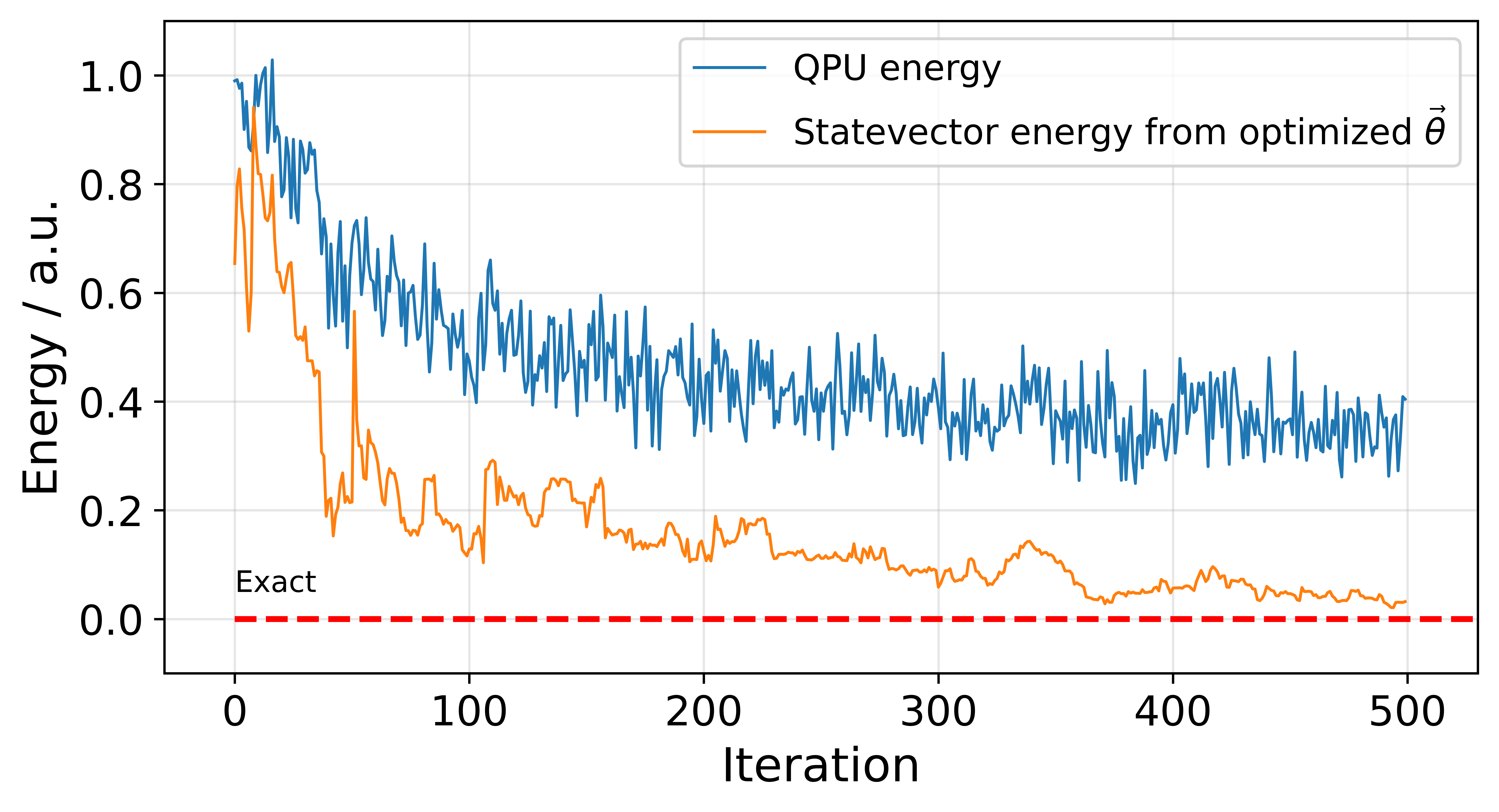}
\caption{VQE optimization for the first torus ground-state candidate. The blue curve shows the energy estimated on the quantum processor, while the cyan curve shows the noiseless statevector energy obtained by evaluating the same variational parameters. The exact zero-energy ground state is also plotted as a dashed red line.}
\label{fig:GS_Torus_Optimization}

\end{figure}

Fig.~\ref{fig:GS_Torus_Optimization} shows the VQE optimization process for the first torus ground-state candidate. During the optimization, the energy measured on the quantum processor remains substantially above the exact zero-energy value and exhibits sizable fluctuations. Taken alone, this noisy hardware signal would suggest a rather poor convergence. 
This trend is a consequence of the SPSA optimization procedure. During the optimization, the stored QPU energy is not an independent evaluation at the central parameter vector, but it is constructed from the perturbed parameter evaluations used internally by classical optimizer (see Supplementary Material for further details).
However, evaluating the same parameter vectors obtained in the real quantum processor with a noiseless statevector simulation reveals that the algorithm properly explores the phase space being able to reach the ground state configuration close the exact solution despite the noise affecting the QPU energy estimates.
The statevector curve is therefore not only a comparison with an ideal calculation, but also an a posteriori diagnostic of the quality of the parameters learned during the noisy optimization. In the Supplemental Material, we further analyze this point by computing the overlap between the states obtained from the optimized QPU parameters and the corresponding exact low-energy states.

\begin{figure}[t]
\centering
\includegraphics[width=\columnwidth]{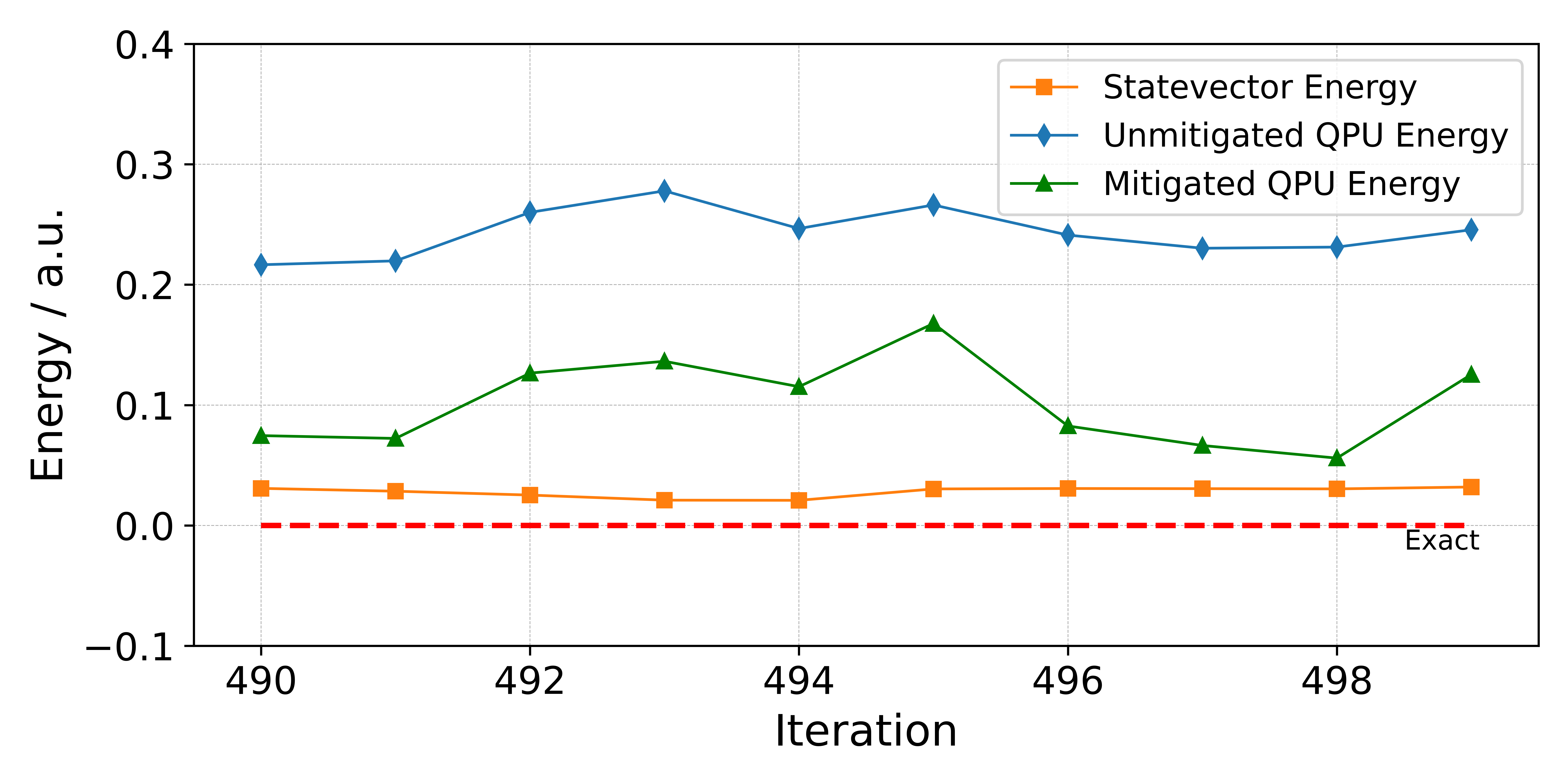}
\caption{
Energy estimates for the last ten parameter vectors of the first torus ground-state optimization. The black squares show the noiseless statevector energies, the pink diamonds show the unmitigated QPU estimates, and the purple triangles show the mitigated results. The dashed line marks the exact zero-energy value. 
}
\label{fig:torus_gs1_zne}
\end{figure}

In order to evaluate the intrinsic error of the real quantum processor, we have recomputed the energies of the last ten variational parameter vectors using noiseless statevector simulation, direct QPU execution, and error mitigation techniques (Zero-noise extrapolation, ZNE, and Twirled Readout Error eXtinction, TREX) \cite{Temme2017, van_den_Berg_2022_TREX, Zepe}. 
The results are shown in Fig.~\ref{fig:torus_gs1_zne}.
Before hardware execution, we also performed a real-time benchmarking step using the corresponding Qiskit tool \cite{IBMRealtimeBenchmarking}, in order to improve qubit selection during the transpilation process. The unmitigated QPU estimates remain systematically shifted upward with respect to the statevector energies, reflecting the effect of hardware noise. ZNE is a low-overhead error-mitigation technique in which the same observable is estimated at different effective noise levels and extrapolated to the zero-noise limit. As we pointed out before, ZNE was combined with TREX, a readout-mitigation protocol that reduces biases associated with measurement errors. Therefore, the mitigated values reported here should be understood as the result of a lightweight mitigation stack: ZNE reduces the effect of gate noise, while TREX mitigates readout errors. This combined protocol partially reduces the systematic upward shift of the QPU energies and brings the estimates closer to the exact zero-energy value, while remaining simple and experimentally low-overhead. The fact that the mitigated values already approach the exact result with such a lightweight mitigation strategy suggests that more advanced, but also more demanding, error-suppression or error-mitigation techniques could further reduce the residual discrepancy.


\begin{figure}[t]
\centering
\includegraphics[width=\linewidth]{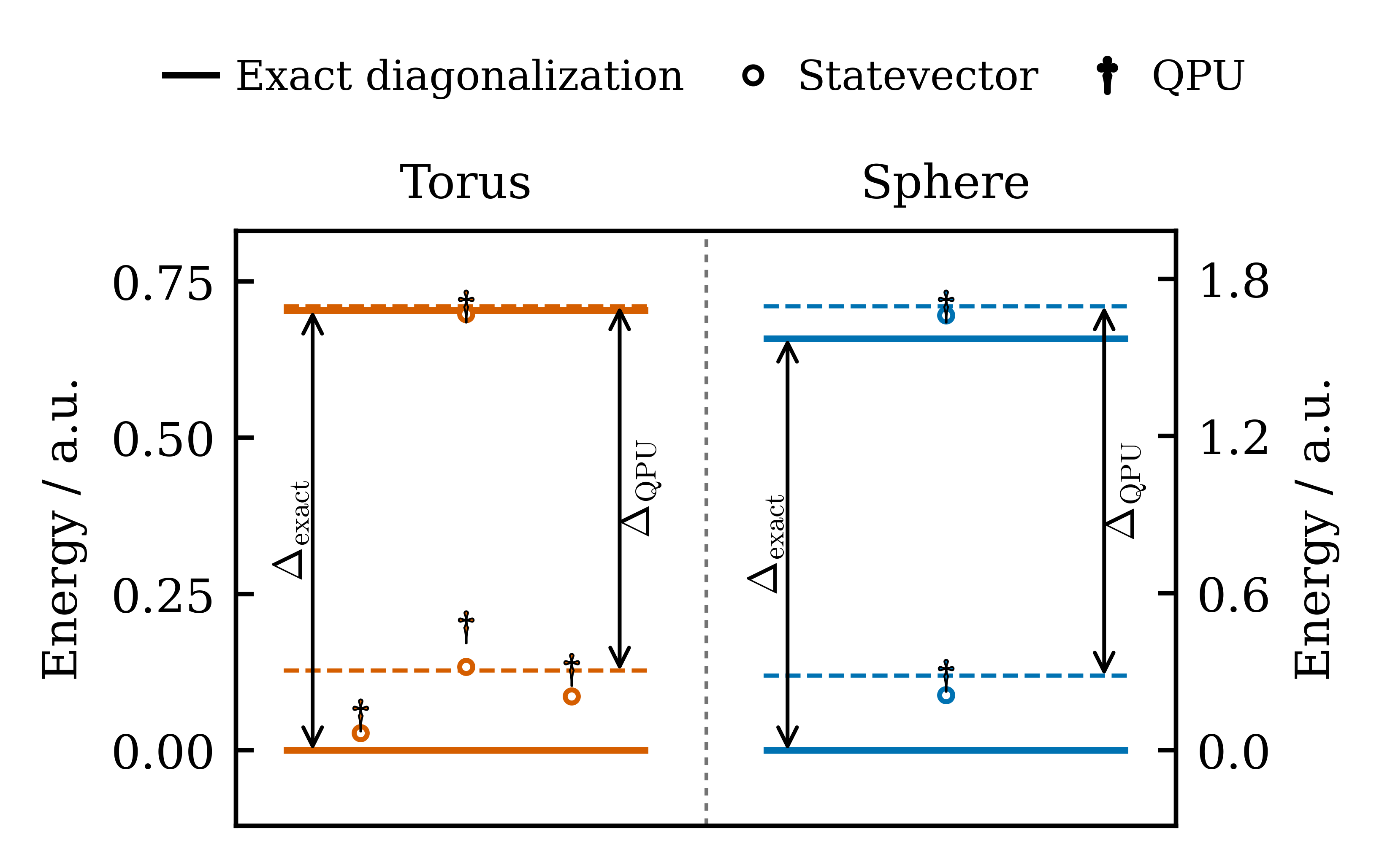}
\caption{
Summary of the energy estimates for the torus and sphere geometries. Left panel: results for the torus, including the three states of the $\nu=1/3$ ground-state manifold and the first excited state. Right panel: results for the sphere, including the unique Laughlin ground state and the first excited state. In each case, exact-diagonalization values are shown as reference levels, and statevector and mitigated results are compared for the optimized variational states. For the torus, the dashed horizontal line denotes the mean QPU energy of the three ground-state estimates; this reference makes explicit the shift in the reconstructed excitation gap due to the hardware noise present in the optimization. In both cases, the plotted values correspond to the mean obtained from the final ten parameter vectors of the corresponding optimization run.
}
\label{fig:torus_summary}
\end{figure}

We then repeated the same workflow used in VQE to the VQD algorithm to reconstruct the remaining two states of the torus ground-state manifold and the first excited state. That is, we optimized the parameter vector for the next states on the real quantum hardware, and then we reevaluate the last ten steps in order to obtain the noiseless statevector energies, raw QPU estimates and the mitigated results.
The resulting energies are summarized in Fig.~\ref{fig:torus_summary} and Table~\ref{tab:torus_results}. 
Performing the parameters obtained in the real hardware at noiseless statevector level, the three variational ground-state candidates remain close to the exact zero-energy manifold, while the first excited state is obtained with an energy very close to the exact-diagonalization value.
The QPU energies are systematically shifted upward, as expected from hardware noise, but the mitigation techniques reduce this bias in all cases. 
For the three ground-state candidates, the best mitigated energies remain clustered close to the exact ground-state manifold, with values $0.056$, $0.197$, and $0.129$. For the first excited state, the best mitigated estimate is $0.710$, in very good agreement with the exact value $0.703$. Thus, even with a low-overhead mitigation strategy, the variational workflow captures both the low-energy clustering of the torus ground-state manifold and the energy scale of the first excitation. This agreement is particularly nontrivial because the Hamiltonian expectation value measured on the quantum processor is reconstructed from a Pauli-string decomposition. 
The accumulated Pauli-coefficient norm sets an error-amplification scale that is substantially larger than the physical energy differences shown in the spectrum, over one order of magnitude in the present case. 
Therefore, the residual deviations of the ZNE estimates from exact diagonalization should be interpreted in the context of this amplified measurement sensitivity. 
A detailed discussion of this error scale is given in the Supplemental Material.

To further assess the quality of the variational states beyond their energies, we analyze their subspace containment with respect to the exact-diagonalization eigenspaces. This diagnostic is computed from the noiseless statevector states obtained by evaluating the QPU-optimized parameter vectors. It therefore probes the ideal wave function encoded by the hardware-optimized circuit, independently of sampling noise and hardware errors. The results are shown in Table~\ref{tab:torus_subspace_containment}.

The three VQD ground-state candidates have large projection weights onto the exact three-dimensional ground-state manifold, with containments $0.970$, $0.883$, and $0.901$, respectively. This confirms that the sequential VQD procedure does not merely produce low energies, but reconstructs states that lie predominantly inside the topological Laughlin ground-state subspace. The last row of Table~\ref{tab:torus_subspace_containment} further shows that the three variational ground-state candidates considered together provide a good approximation to the full exact ground-state manifold, with an overlap of $0.918$.

The optimized first excited state requires a more careful interpretation. Its energy is very close to the exact first-excited value, as shown in Table~\ref{tab:torus_results}, but its containment in the exact first-excited subspace is $0.608$. The remaining weight is distributed mainly between the ground-state manifold, with weight $0.152$, and the second excited subspace, with weight $0.227$. Thus, the energy agreement should not be interpreted as an equally accurate reconstruction of the first-excited wave function. Rather, it reflects that the variational state has substantial overlap with the target excited subspace, together with residual leakage into neighboring low-energy sectors.

In addition to the energy, we have resolved the optimized states according to the magnetic momentum sector $K$. This quantity is obtained directly from samples of the quantum circuit in the occupation number basis. For each measured bit string, interpreted as an occupation configuration $|n_0n_1\cdots n_{N_\phi-1}\rangle$, we have computed $K$ as defined in Eq.~\eqref{eq:torus_K}. The distribution of measured bit strings therefore gives a distribution over magnetic momentum sectors. We assign to each optimized state the sector weight distribution when more than one sector has appreciable probability. This provides a torus-specific symmetry diagnostic, complementary to the energy and to the overlap analysis reported in the Supplemental Material.

\begin{table*}[t]
\squeezetable \begin{ruledtabular} \begin{tabular}{llcccccc} Observable & State & Exact diag. & Statevector avg. & Unmitigated avg. & Best mitigated & Mitigated avg. & \(W_{\mathrm{tar}}\) (\%) \\ \hline \(E\) & GS$_1$ & \(0\) & \(0.0280 \pm 0.0039\) & \(0.244 \pm 0.019\) & \(0.056\) & \(0.102 \pm 0.035\) & \(88.16\) \\ \(E\) & GS$_2$ & \(0\) & \(0.13336 \pm 0.00062\) & \(0.414 \pm 0.023\) & \(0.197\) & \(0.242 \pm 0.034\) & \(89.38\) \\ \(E\) & GS$_3$ & \(0\) & \(0.08633 \pm 0.00051\) & \(0.394 \pm 0.026\) & \(0.129\) & \(0.180 \pm 0.036\) & \(84.64\) \\ \(E\) & 1st excited & \(0.703\) & \(0.698 \pm 0.014\) & \(0.907 \pm 0.027\) & \(0.710\) & \(0.774 \pm 0.043\) & \(75.88\) \\ \end{tabular} \end{ruledtabular} 
\caption{Results for the torus geometry. Statevector and unmitigated QPU values are reported as the mean and standard deviation obtained from the final ten parameter vectors of each optimization process. ZNE and TREX were likewise applied to each of these ten parameter vectors. The column ``Best mitigated'' reports the mitigated estimate closest to the corresponding statevector value, while ``Mitigated avg.'' gives the mean value and standard deviation over the ten mitigated results. The target momentum-subspace weight is defined as \(W_{\mathrm{tar}}=\sum_{K\in\mathcal K_{\mathrm{tar}}}p(K)\), with \(\mathcal K_{\mathrm{tar}}=\{1,3,5\}\) for the ground-state manifold and \(\mathcal K_{\mathrm{tar}}=\{0,2,4\}\) for the first excited state.}
\label{tab:torus_results}
\end{table*}

Since the three Laughlin ground states are degenerate on the torus, $GS_1$, $GS_2$, and $GS_3$ should be understood as the three states obtained sequentially by VQD within the same zero-energy manifold, rather than as energetically distinct levels. In Table~\ref{tab:torus_results}, we report the total weight in the target momentum subspace, $W_{\mathrm{tar}}=\sum_{K\in\mathcal K_{\mathrm{tar}}}p(K)$. For the ground-state manifold, the target subspace is $\mathcal K_{\mathrm{tar}}={1,3,5}$, while for the first excited state it is $\mathcal K_{\mathrm{tar}}={0,2,4}$. The three VQD ground-state candidates have large target-subspace weights, $88.16\%$, $89.38\%$, and $84.64\%$, showing that the optimized circuits concentrate most of the probability in the expected magnetic-momentum sectors of the Laughlin manifold. The first excited state has a lower, but still dominant, target-subspace weight of $75.88\%$, consistent with the greater sensitivity of excited-state preparation to variational imperfections and hardware noise. The full set of energy estimates and momentum-sector weights is reported in Table~\ref{tab:torus_results}.

\begin{table*}[t]
\squeezetable
\begin{ruledtabular}
\begin{tabular*}{\textwidth}{@{\extracolsep{\fill}}lcccc}
State & GS subspace & 1st exc. subspace & 2nd exc. subspace & 3rd exc. subspace \\
\hline
GS$_1$ & $0.970$ & $0.0046$ & $0.0186$ & $0.0066$ \\
GS$_2$ & $0.883$ & $0.0085$ & $0.0446$ & $0.0637$ \\
GS$_3$ & $0.901$ & $0.0412$ & $0.0559$ & $0.0016$ \\
1st excited & $0.152$ & $0.608$ & $0.227$ & $0.0126$ \\
\hline
Variational GS subspace
& \multicolumn{4}{c}{$0.918$ overlap with the exact GS manifold} \\
\end{tabular*}
\end{ruledtabular}
\caption{
Subspace containment analysis for the torus geometry. For each optimized state,
we report the projection weight onto the exact ED eigenspaces. The GS subspace
corresponds to the threefold zero-energy Laughlin manifold. The last row reports
the overlap between the three-dimensional variational GS subspace and the exact
GS manifold.
}
\label{tab:torus_subspace_containment}
\end{table*}

\subsection{Sphere geometry}
\label{subsec:results_sphere}

We now turn to the spherical geometry. In contrast to the torus, the $\nu=1/3$ Laughlin state on the sphere presents a unique zero-energy ground state for $N_L=7$ $(N_e=3)$. We therefore use VQE to target the ground state and VQD to obtain the first excited state. The results are summarized in Fig.~\ref{fig:torus_summary} and Table~\ref{tab:sphere_results}. 
Performing a noiseless statevector evaluation with the parameters obtained from the real quantum hardware measurements demonstrates that both states are recovered with energies very close to the exact-diagonalization benchmarks. The QPU energies again show a systematic upward shift due to hardware noise, while ZNE and TREX partially correct this bias and improves the agreement with the exact spectrum.

\begin{table*}[t]
\squeezetable
\begin{ruledtabular}
\begin{tabular}{llccccc}
Observable & State & Exact diag. & Statevector avg. & Unmitigated avg. & Best mitigated & Mitigated avg. \\
\hline
\(E\) 
& GS 
& \(0\) 
& \(0.2115 \pm 0.0070\) 
& \(1.024 \pm 0.051\) 
& \(0.285\) 
& \(0.49 \pm 0.11\) \\
                
\(E\) 
& 1st excited 
& \(1.5114\) 
& \(1.6604 \pm 0.0010\) 
& \(2.161 \pm 0.053\) 
& \(1.695\) 
& \(1.87 \pm 0.11\) \\
                
\(\langle \hat L^2\rangle\) 
& GS 
& \(0\) 
& \(1.306 \pm 0.029\) 
& \(5.51 \pm 0.16\) 
& \(2.22\) 
& \(2.83 \pm 0.31\) \\
                
\(\langle \hat L^2\rangle\) 
& 1st excited 
& \(12\) 
& \(12.3032 \pm 0.0093\) 
& \(13.96 \pm 0.19\) 
& \(12.53\) 
& \(12.88 \pm 0.22\) \\
\end{tabular}
\end{ruledtabular}
\caption{Results for the spherical geometry. Statevector and unmitigated QPU values are reported as the mean and standard deviation obtained from the final ten parameter vectors of each optimization process. ZNE and TREX were likewise applied to each of these ten parameter vectors. The column ``Best mitigated'' reports the mitigated estimate closest to the corresponding statevector value, while ``Mitigated avg.'' gives the mean value and standard deviation over the ten mitigated results.}
\label{tab:sphere_results}
\end{table*}

For the ground state, the best mitigated result gives an energy of $0.285$, compared with the exact value $E=0$. For the first excited state, the best ZNE mitigated energy value is $1.695$, while the exact diagonalization one is $1.5114$.
Like in the torus case, these deviations should be interpreted in the context of the Pauli-string reconstruction of the Hamiltonian expectation value, where measurement errors are amplified by the coefficient norm of the qubit Hamiltonian. Despite this sensitivity, the mitigated results reproduce the qualitative structure of the spectrum, namely a low-energy Laughlin ground state separated from the first excited state by a finite gap (see Fig. ~\ref{fig:torus_summary} .

We also perform the same subspace-containment analysis for the spherical geometry. The results are reported in Table~\ref{tab:sphere_subspace_containment}. Since the relevant spherical eigenstates are nondegenerate in the cases considered here, the containment reduces to the squared overlap with the corresponding exact ED eigenstate. The optimized ground state has a dominant projection onto the exact Laughlin ground-state subspace, with containment $0.907$. Similarly, the optimized first excited state has a dominant projection onto the exact first-excited subspace, with containment $0.906$. The remaining weight is distributed among higher excited subspaces and remains comparatively small. These results confirm that the QPU-optimized parameters define, in the noiseless statevector limit, variational states with the correct wave-function character, not only reasonable energy estimates.

\begin{table*}[t]
\squeezetable
\begin{ruledtabular}
\begin{tabular*}{\textwidth}{@{\extracolsep{\fill}}lccccc}
State & GS subspace & 1st exc. subspace & 2nd exc. subspace & 3rd exc. subspace & 4th exc. subspace \\
\hline
GS
& $0.907$
& $0.0274$
& $0.0361$
& $0.0220$
& $0.00740$ \\
1st excited
& $0.0133$
& $0.906$
& $0.0324$
& $0.0357$
& $0.0127$ \\
\end{tabular*}
\end{ruledtabular}
\caption{
Subspace containment analysis for the spherical geometry. For each optimized state,
we report the projection weight onto the corresponding exact ED eigenspaces.
}
\label{tab:sphere_subspace_containment}
\end{table*}

 The angular-momentum diagnostic in Table~\ref{tab:sphere_results} gives a more detailed picture of the quality of the optimized states. 
 For the ground state, the exact value is $\langle \hat{L}^2\rangle=0$. 
 The statevector result, $\langle \hat{L}^2\rangle=1.306\pm0.029$, shows that the variational state is close to the low-energy Laughlin sector but still contains a residual admixture of higher-angular-momentum components. On the quantum processor, the unmitigated value increases to $5.51\pm0.16$, reflecting the strong sensitivity of $\hat{L}^2$ to hardware noise. ZNE adn TREX improve this estimate, giving a best value of $2.22$ and an average value of $2.83\pm0.31$. The first excited state is reproduced more accurately in angular momentum: the exact value is $\langle \hat{L}^2\rangle=12$, while the statevector calculation gives $12.3032\pm0.0093$. The unmitigated QPU value is shifted to $13.96\pm0.19$, but ZNE brings it close to the correct sector, with a best value of $12.53$ and an average value of $12.88\pm0.22$. These results show that the variational workflow captures not only the energy ordering, but also the approximate angular-momentum character of the spherical low-energy states.

\section{Conclusions}
\label{sec:conclusions}    
We have studied the preparation and characterization of FQHE states using hybrid variational quantum algorithms on both noiseless simulators and real quantum hardware. Focusing on the fermionic $\nu=1/3$ Laughlin phase described by the $V_1$ Haldane pseudopotential, we formulated the problem in second quantization after projection onto the LLL and mapped the resulting Hamiltonians to qubits through the Jordan--Wigner transformation. The use of particle-number-preserving variational circuits, combined with VQE and VQD, allowed us to target not only individual low-energy states, but also degenerate manifolds.

The torus geometry provides the central benchmark of this work. In contrast to the sphere, the torus has no shift and supports the characteristic threefold topological ground-state degeneracy of the $\nu=1/3$ Laughlin phase. We showed that VQD can sequentially reconstruct the degenerated ground state manifold and first excited state of the torus. The parameters obtained through the quantum hardware routine evaluated at the statevector level produce energies which reproduce the exact diagonalization spectrum with good accuracy. On the quantum processor, the unmitigated energies are systematically shifted upward by hardware noise, but ZNE and TREX substantially reduces this bias. The first excited state energy is particularly well reproduced after mitigation, while the three ground-state candidates remain clustered close to the zero-energy manifold. The subspace-containment analysis further shows that the QPU-optimized parameters reconstruct the topological ground-state manifold at the wave-function level, not only at the level of energy estimates.

Beyond the energy, we used the magnetic momentum sector $K$ as a torus-specific symmetry diagnostic. By sampling the optimized circuits in the occupation number basis and computing $K$ for each bit string, we extracted the momentum-sector distributions of the prepared states. The three VQD ground-state candidates show large weight in the target momentum subspace associated with the Laughlin manifold, confirming that the algorithm captures not only low energies, but also the expected magnetic-translation structure. This is important because the torus ground states are exactly degenerate: energy alone is not sufficient to identify whether the correct topological manifold has been reconstructed.

The spherical geometry provides a complementary benchmark in which the Laughlin state is a unique angular-momentum singlet. In this case, VQE targets the zero-energy ground state and VQD gives access to the first excited state. The energy estimates reproduce qualitatively the energy spectrum of the structure, namely a low-energy Laughlin ground state separated from the first excitation by a finite gap. We also evaluated $\langle \hat{L}^2\rangle$ as a symmetry diagnostic. 
The parameters obtained through the quantum hardware routine evaluated at the statevector level produce results that are close to the expected angular-momentum sectors, and and TREX improve the corresponding QPU estimates, especially for the first excited state. This confirms that the variational workflow captures part of the angular-momentum structure of the spherical low-energy spectrum.

An important aspect of the results is that the agreement with exact diagonalization is achieved using a relatively low-overhead mitigation strategy. The Hamiltonian expectation values are reconstructed from Pauli-string decompositions, so hardware errors are amplified by the coefficient norm of the qubit Hamiltonian. In the present calculations this error-amplification scale is substantially larger than the physical energy differences of interest, as discussed in the Supplementary Material. The fact that ZNE and TREX already recover the main spectral features under these conditions suggests that more advanced error-suppression and error-mitigation techniques could further improve the quantitative accuracy.

The broader significance of these results is that they establish a controlled quantum-computing workflow for strongly correlated topological matter beyond simple spin or molecular benchmarks. The continuum fractional quantum Hall problem provides an ideal starting point because it combines an exactly known topological phase, a compact second-quantized formulation, nontrivial degeneracies, and sharp symmetry diagnostics. Having demonstrated that real quantum processors are in the path to recover the physics behind structures, the natural next step is to move from benchmark parent Hamiltonians to realistic two-dimensional Chern-band problems, where no exact Laughlin wave function is available and classical exact diagonalization becomes rapidly prohibitive. Fractional Chern insulators, moiré flat bands, and other interacting topological materials provide precisely this setting: they retain the same physical ingredients of topology, interactions, and fractionalization, but add the microscopic complexity of real lattice Hamiltonians. In this sense, the present work is not only a demonstration of fractional quantum Hall state preparation, but a platform for approaching quantum utility in realistic correlated topological phases, where quantum hardware may eventually access low-energy manifolds, topological quantum numbers, and many-body observables beyond the reach of classical methods.

\begin{acknowledgments}

F. G. and J. B. want to thank V\~{o} Ti\'{e}n Phong for fruitful discussions. We acknowledge support from the Basq initiative. 
This work has been possible thanks to the quantum computational resources provided by the BasQ Strategy under the collaboration agreement between Ikerbasque Foundation and the University of the Basque Country (UPV/EHU), the Donostia International Physics Centre (DIPC) and the University on Mondragon, on behalf of the Department of Science, Universities and Innovation of the Basque Government. The authors thank for technical and human support provided by Scientific Computing Service of the SGIker (UPV/EHU/FEDER, EU).
F. G. acknowledges support from the ``Severo Ochoa'' Programme for Centres of Excellence in R\&D (CEX2020-001039-S/AEI/10.13039/501100011033) financed by MICIU/AEI/10.13039/501100011033 and from NOVMOMAT, Grant PID2022-142162NB-I00 funded by MCIN/AEI/10.13039/501100011033 and by ``ERDF A way of making Europe''. 
U. A. and R. G-A. acknowledges support from the Gipuzkoa Quantum Programme financed by the Provincial Government of Gipuzkoa with project number 2025-QUAN-000021-01 and 2025-QUAN-000037-01. 
U. A. and R. G-A. acknowledge support from the ELKARTEK programme financed by the Basque Goverment with project number KK-2025/00079.
\end{acknowledgments}

\nocite{Combarro2023ExpectationValues}
\nocite{Zepe}
\nocite{Wilson2020}
\nocite{IBMRealtimeBenchmarking}
\bibliographystyle{apsrev4-2}
\bibliography{Biblio-FQHE}

\end{document}